\documentclass[epj]{webofc}
\usepackage[utf8]{inputenc}
\usepackage[varg]{txfonts}   
\usepackage{booktabs}
\usepackage{xcolor}
\definecolor{darkred}{rgb}{0.4,0.0,0.0}
\definecolor{darkgreen}{rgb}{0.0,0.4,0.0}
\definecolor{darkblue}{rgb}{0.0,0.0,0.4}
\usepackage[bookmarks,linktocpage,colorlinks,
    linkcolor = darkred,
    urlcolor  = darkblue,
    citecolor = darkgreen]{hyperref}
\newcommand{\av}[1]{\left\langle #1 \right\rangle} 
\renewcommand{\bar}[1]{\overline{#1}}%
\newcommand{\mass}{ am_\text{q} }
\newcommand{\fig}{Fig.\xspace}
\newcommand{\tab}{Tab.\xspace}

\newcommand{\suthree}{SU(3)\xspace}
\newcommand{\uone}{U(1)\xspace}
\wocname{EPJ Web of Conferences}
\woctitle{Lattice2017}
\begin{document}
%
\selectlanguage{english}
\title{
Chiral phase transition of three flavor QCD with nonzero magnetic field using standard staggered fermions
}
\author{%
\firstname{Akio} \lastname{Tomiya}\inst{1}
\fnsep\thanks{Speaker, \email{akio.tomiya@mail.ccnu.edu.cn} }
\and
\firstname{Heng-Tong} \lastname{Ding}\inst{1} \and
\firstname{Swagato} \lastname{Mukherjee}\inst{2} \and
\firstname{Christian} \lastname{Schmidt}\inst{3} \and
\firstname{Xiao-Dan}  \lastname{Wang}\inst{1}
}
\institute{%
Key Laboratory of Quark \& Lepton Physics (MOE) and Institute of Particle Physics, Central China Normal University, Wuhan 430079, China
\and
Department of Physics, Brookhaven National Laboratory, Upton, New York 11973-5000
\and
Fakult\"at f\"ur Physik, Universit\"at Bielefeld, D-33615 Bielefeld, Germany 
}
\abstract{%
Lattice simulations for (2+1)-flavor QCD with external magnetic field demonstrated that the quark mass is 
one of the important parameters responsible for the (inverse) magnetic catalysis. 
We discuss the dependences of chiral condensates and susceptibilities, the Polyakov loop on the magnetic field and quark mass in three degenerate flavor QCD. 
The lattice simulations are performed using standard staggered fermions and the plaquette 
action with spatial sizes $N_\sigma$ = 16 and 24 and a fixed temporal size $N_\tau$ = 4. 
The value of the quark masses are chosen such
that the system undergoes a first order chiral phase transition and crossover with zero magnetic field. 
We find that in light mass regime, the quark chiral condensate undergoes magnetic catalysis in the whole temperature region
and the phase transition tend to become stronger as the magnetic field increases.
In crossover regime, deconfinement transition temperature is shifted by the magnetic field when quark mass $ma$ is less than $0.4$.
The lattice cutoff effects are also discussed.
}
\maketitle
\section{Introduction}\label{intro}
{A}
{strong magnetic field is expected to be produced in the early stage of the peripheral heavy ion collisions \cite{Kharzeev:2007jp}.
This interesting phenomenon drives a lot of theoretical studies since the external magnetic field affects the system through the electric charge of quarks especially phase and phase transition of QCD matter \cite{Endrodi:2014vza,Kharzeev:2015znc}.}

From a view point of the field theory, the external magnetic field $B$ breaks several symmetries:
1) Rotational symmetry,
2) Time reversal symmetry and
3) Isospin symmetry.
These breakings allow to mix channels which are prohibited by the symmetries of the original theory.
For example,
1) allows to mix (axial)vector and (pseudo)scalar channels.
Breaking of 3)  can be obtained by introducing an isospin chemical potential $\mu_\text{iso}$ but
the external magnetic field also breaks 1)  and 2) ,
thus the phase diagram is expected to be different from a $\mu_\text{iso}\neq 0$ system.

Recently QCD phase structure with magnetic field has been intensively studied on the lattice \cite{Endrodi:2014vza}.
Earlier studies, that employed fourth rooted standard staggered fermions with $N_f = 2$ found that the critical temperature $T_c$ increases as a function of magnitude of the magnetic field $B$ and the strength of the phase transition becomes stronger with larger magnetic field \cite{d2010qcd}. Afterwards by using 
stout staggered fermions with a physical pion mass a totally different result was found. There the inverse catalysis and the decreasing of $T_c$ with $B$ was observed \cite{bali2012qcd, bali2012qcd2}. In the simulation with standard staggered fermions the root-mean-square pion mass is much larger than the physical pion mass and this could be a cause of the observation of magnetic catalysis \cite{Bruckmann:2013oba}.{}

\begin{figure}[htbp]
        \begin{center}
              \begin{minipage}{0.35\hsize }
                \begin{center}
                        \includegraphics[width=\hsize]{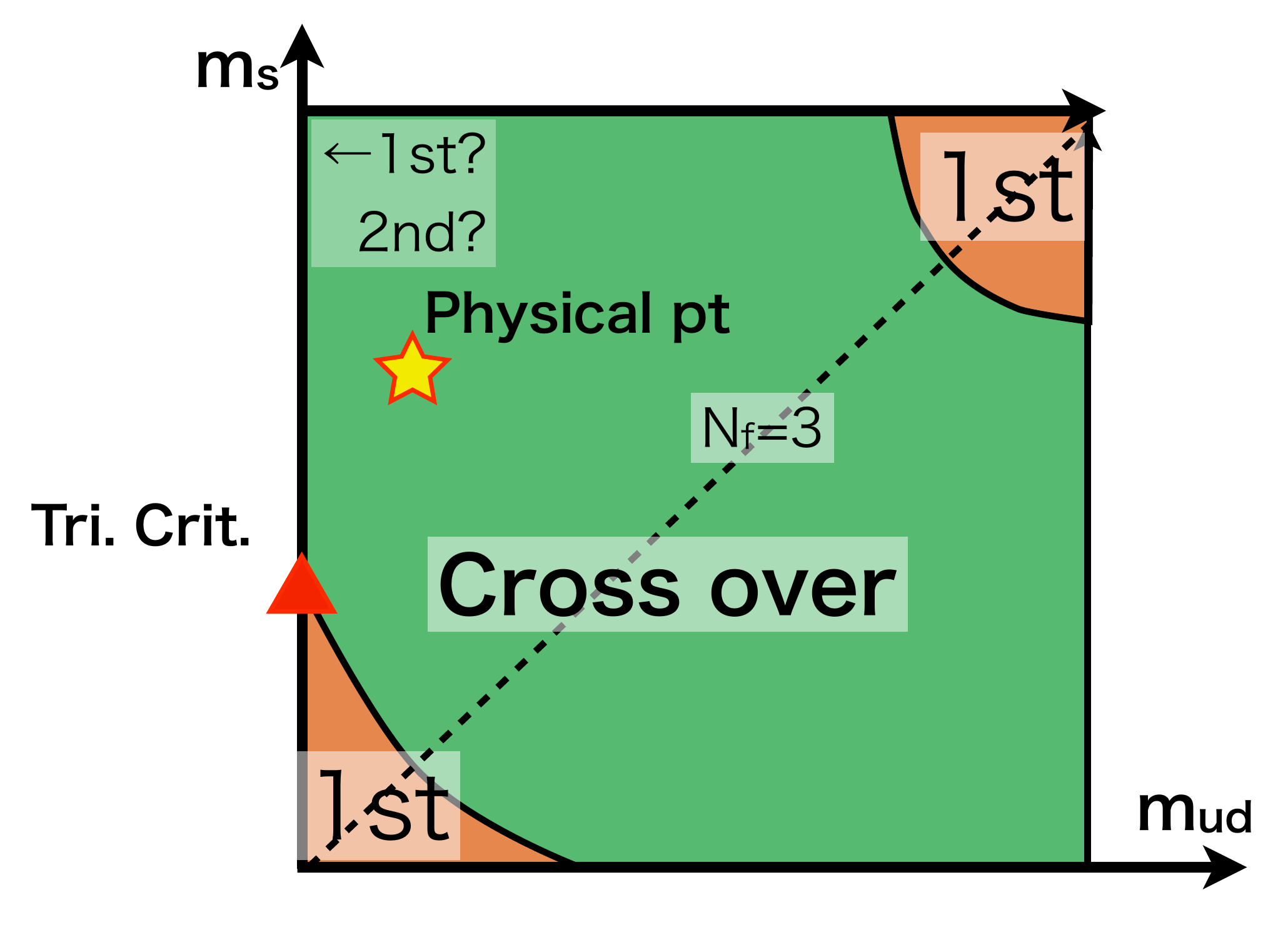}
                \end{center}
              \end{minipage}
              \begin{minipage}{0.315\hsize }
                \begin{center}
                        \includegraphics[width=\hsize]{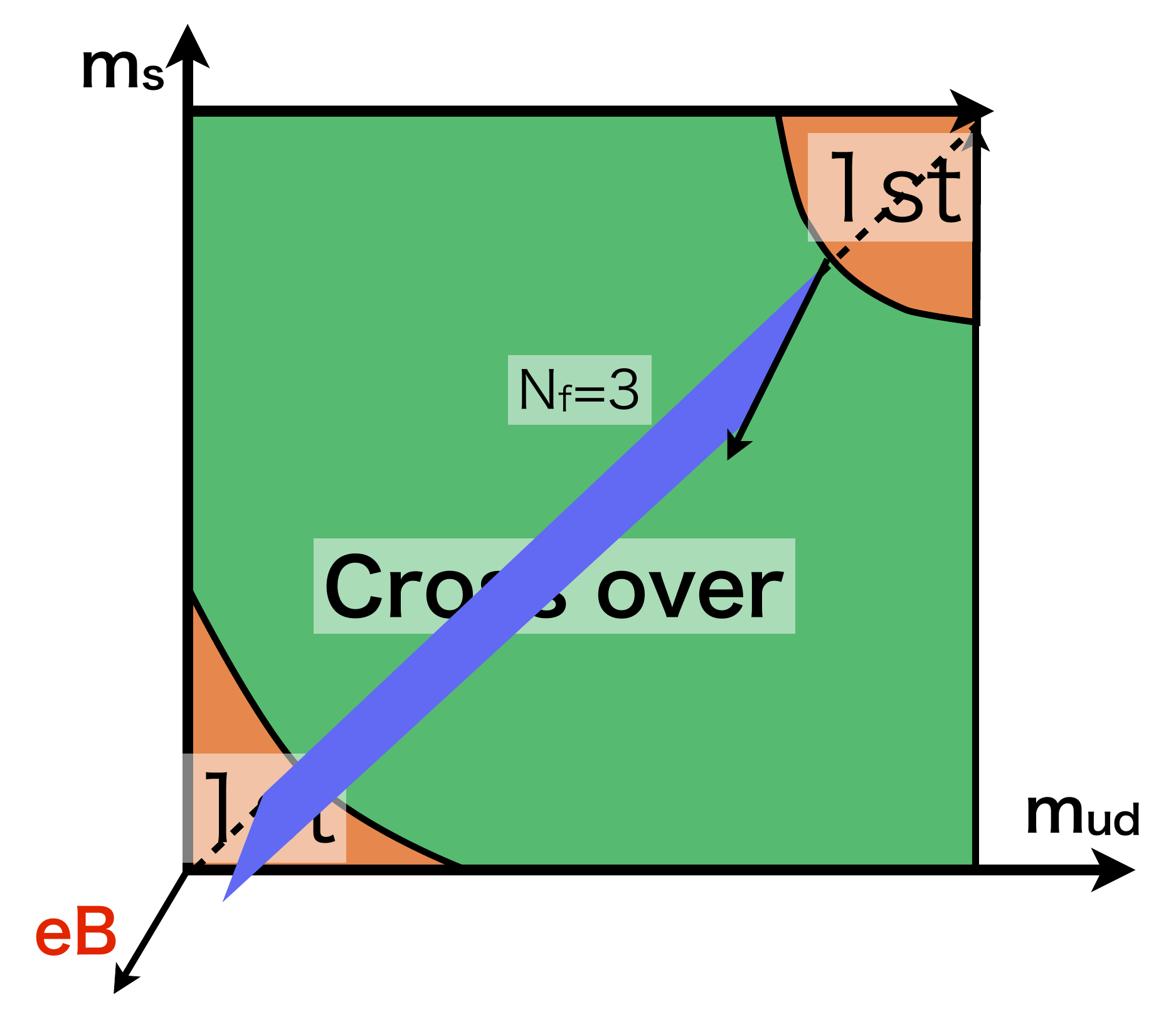}
                \end{center}
              \end{minipage}
        \end{center}
\caption{
Conventional Colombia plot \cite{Brown:1990ev} (left) and the one with the external magnetic field (right).
\label{fig:Colombia_plot} }
\end{figure}

In this proceedings, we intend to study the influence of the value of quark mass to the behavior of $T_c$ as a function of the magnetic field $B$, namely we investigate $\beta_\text{cri}(B,m)$ with $N_f=3$ QCD, which corresponds to a diagonal line for the extended Columbia plot (Right panel in Fig. \ref{fig:Colombia_plot}). As a starting point of this project we will use the standard staggered fermions in our simulation. The numerical setups will be presented in the next section.

%
%

\section{Setup}
\label{sec:setup}
Here we introduce our numerical setup.
We employ mass degenerated three flavor standard staggered fermions, 
$am_\text{ud} = am_\text{s} \equiv \mass$, with the plaquette gauge action.
Parameters for light mass regime are taken from \cite{Smith:2011pm},
with which QCD has a first order phase transition at vanishing magnetic field $B=0$. 
This simulation is performed with forth rooting technique as in previous studies
and rational hybrid Monte-Carlo algorithm. 

The external \uone magnetic field is implemented in the following way.
The magnetic field only couples to quarks thus implementation is done just by changing \suthree links $U_\mu$ to
$u_\mu U_\mu$. Here $u_\mu$ represents \uone links
which contribute to the Dirac operator.
Finiteness of lattice size introduces an infrared cutoff to the \uone field \cite{al2009discrete}.
Let us denote the lattice size $(N_x,\; N_y,\; N_z,\; N_t)$ and coordinate as $n_\mu=0,\cdots, N_\mu-1$ ($\mu = x,\; y,\; z,\; t$).
The external magnetic field in $z$ direction $\vec{B}=(0,0,B)$ is described by the link variable $u_\mu(n)$ of the \uone field and 
$u_\mu(n)$ is expressed as follows,
\begin{align}
u_x(n_x,n_y,n_z,n_t)&=
\begin{cases}
\exp[-iq {B} N_x n_y] \;\;&(n_x= N_x-1)\\
1 \;\;&(\text{Otherwise})\\
\end{cases}\notag\\
u_y(n_x,n_y,n_z,n_t)&=\exp[iq {B}  n_x],\label{eq:def_mag_u} \\
u_z(n_x,n_y,n_z,n_t)&=u_t(n_x,n_y,n_z,n_t)=1.\notag
\end{align}
where $q$ is the electric charge of each quark. 
%
One-valuedness of the one particle wave function requires the Dirac quantization 
to be similar as in superconductors,
\begin{align}
q\hat{B} = \frac{2\pi N_b}{N_x N_y} \label{eq:mag_to_Nb},
\end{align}
where $N_b \in { \boldsymbol Z}$ is the number of magnetic flux through unit area for $x$-$y$ plane
and $\hat{B} \equiv a^2 B$.
The ultraviolet cutoff $a$ introduces also a periodicity of the magnetic field along with $N_b$.
Namely, a range
$0\leq N_b < {N_x N_y}/{4}$
represents an independent magnitude of the magnetic field $B$ similar to momentum in the Brillouin zone for crystals.

In order to examine effects on the phase transition coming from mass and the external field,
we introduce a ``threshold mass''.
The continuum Dirac operator with a \suthree-valued gluon field and external magnetic field is ${M}_\text{con.} = \gamma_\mu (D_\mu -i q a_\mu ) + m $.
By inserting the external magnetic field $a_\mu=\; (0,Bx,0,0)$ in the Landau gauge we obtain, 
\begin{align}
{M}_\text{con.} ^\dagger {M}_\text{con.}
&= D_\mu^2 + (qxB)^2+m^2.
\end{align}
This suggests that the $qB$ term plays a similar role as the mass $m$ in this gauge.
We introduce the ``threshold mass'' with $N_b$ by a dimensional analysis, 
\begin{align}
a\sqrt{eB} = \sqrt{\frac{2\pi N_b}{N_x N_y |q|} } \equiv a m_{N_b}^\text{th}.
\end{align}
This mass is considered to be a threshold for magnetic field dominated regime to a mass dominated regime.
We take $q=2/3$ in our study.
By setting $a m_{N_b}^\text{th}$, and if taking $N_b$ larger than that value,
the system thus may be considered to be dominated by the magnetic field rather than the quark mass.
Details of our parameters are summarized in \tab\ref{tab:parameters}.

%
\begin{table}[htb]
\center
\begin{tabular}{c|c|c|c|c|c} 
$N_\sigma^3\times N_\tau$ & $\mass$ & $\beta$ range & $N_b$ & \# Conf. & Note\\ \hline\hline
$24^3\times 4$ & $0.024$ & 5.128 -- 5.160 & 0 -- 56 &$O(2000)$& First order for $N_b=0$ \\\hline
$16^3\times 4$ & $0.028$ & 5.130 -- 5.170 & 0 -- 56 & $O(1500)$& Crossover for $N_b=0$ \\\hline
$16^3\times 4$ & $0.2    $ & 5.10 -- 5.65  & 0 -- 56 &$ O(500)$& Around $a m_{N_b=1}^\text{th}$ for up quark \\\hline
$16^3\times 4$ & $0.4    $ & 5.35 -- 5.65 & 0 -- 56 & $O(500)$& Around $a m_{N_b=4}^\text{th}$ for up quark  \\\hline
$16^3\times 4$ & $0.8    $ & 5.35 -- 5.85 & 0 -- 56 & $O(700)$& Around $a m_{N_b=14}^\text{th}$ for up quark
\end{tabular}
\caption{
Summary of our numerical setup.
The  order of phase transition for $N_b=0$ is determined in \cite{Smith:2011pm}.
\label{tab:parameters}
}
\end{table}

\section{Results}
\label{sec:results}
Here we summarize our preliminary results, which are listed in \tab\ref{tab:results}.
We measure the chiral condensate and the Polyakov loop and their susceptibilities.
The critical temperature (critical $\beta$) for the chiral phase transition is determined by the susceptibility of chiral condensates in the light mass regime.
The confinement/deconfinement transition is determined by one of the Polyakov loop in the heavy mass regime.
In the light mass regime, we calculate the Binder cumulant \cite{binder1981critical} for the chiral condensate
as a function of $\beta$,
\newcommand{\deltapsi}{\delta\bar{\psi}\psi}
\begin{align}
B_4 (\beta) = \frac{ \av {(\deltapsi)^4} }{ \av{(\deltapsi)^2}^2 },
\end{align}
where $\deltapsi = \bar{\psi}\psi - \av{\bar{\psi}\psi} $.
The minimum of the Binder cumulant $B_4$ indicates the order of the phase transition:
$B_4 =3$ corresponds to a crossover,
$B_4 \sim 1.6$ to a second order phase transition with the Ising $Z_2$ universality class,
$B_4 = 1$ to a first order phase transition \cite{Bazavov:2017xul}.
However, $B_4$ contains forth order fluctuations, thus it is difficult to obtain accurate result compared to
the susceptibility or the condensate itself.

\begin{table}[htb]
\center
\begin{tabular}{c|c|c|c|c|c} 
$N_\sigma^3\times N_\tau$ & $\mass$ & $\beta_\text{crit}$  along with $N_b$ & Chiral & Confinement\\ \hline\hline
$24^3\times 4$ & $0.024$ & Increase* & 1st& 1st\\\hline
$16^3\times 4$ & $0.028$ & Increase*  & Crossover to critical$^\dagger$ & Crossover to critical$^\dagger$\\\hline
$16^3\times 4$ & $0.2    $ & Increase**   & Not critical & Crossover-like\\\hline
$16^3\times 4$ & $0.4    $ & Increase**  &  Not critical & Crossover-like\\\hline
$16^3\times 4$ & $0.8    $ & Not clear  &  Not critical & Phase transition-like***
\end{tabular}
\caption{
Summary of our preliminary results.
* $\beta_\text{crit}$ for light mass regime is determined by the up quark condensate
and the Polyakov loop shows similar critical behavior.
** $\beta_\text{crit}$ for heavy mass regime is the Polyakov loop.
*** From shape of the Polyakov loop susceptibility.
$^\dagger$ By the Binder cumulant.
\label{tab:results}
}
\end{table}


%

\begin{figure}[htbp]
        \begin{center}
              \begin{minipage}{0.55\hsize }
                \begin{center}
                        \includegraphics[width=\hsize]{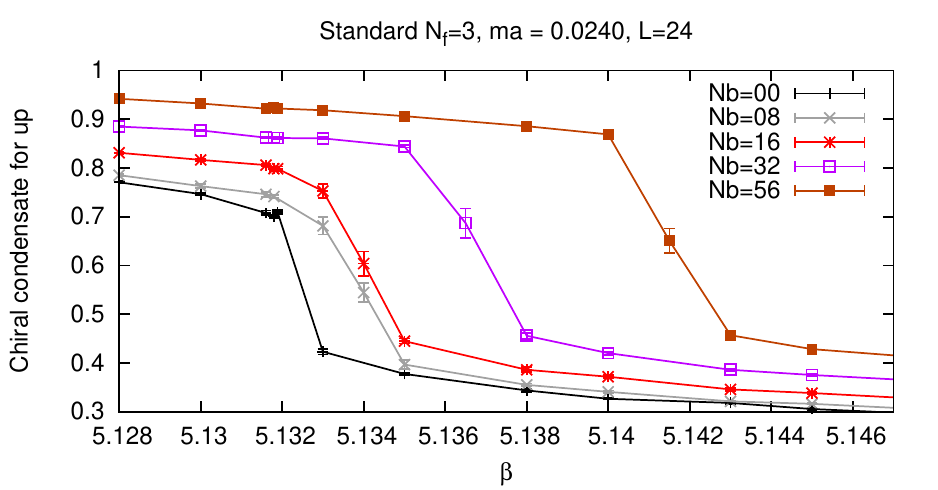}
                \end{center}
              \end{minipage}
              \begin{minipage}{0.55\hsize }
                \begin{center}
                        \includegraphics[width=\hsize]{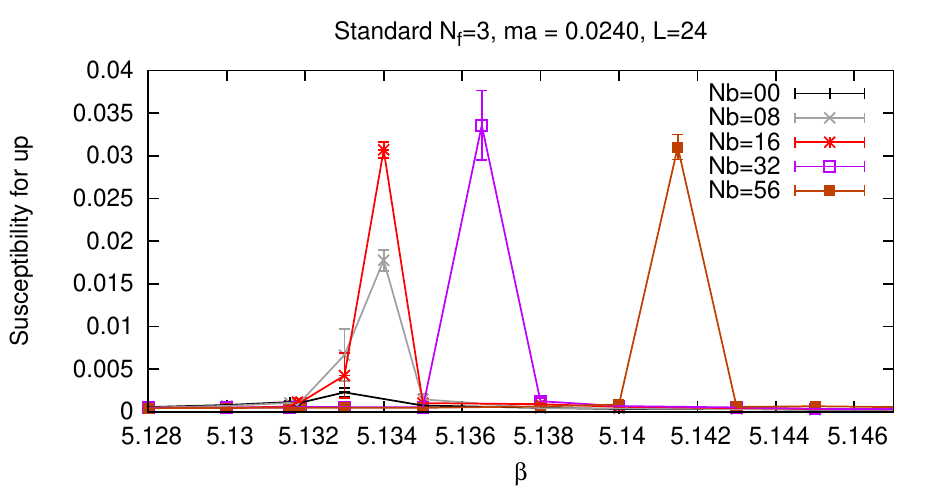}
                \end{center}
              \end{minipage}
                          \begin{minipage}{0.55\hsize }
                \begin{center}
                        \includegraphics[width=\hsize]{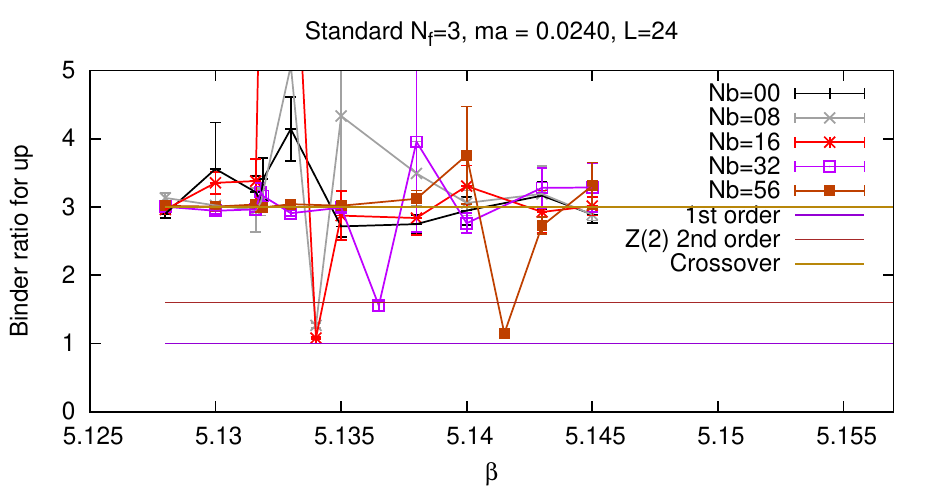}
                \end{center}
              \end{minipage}
        \end{center}
\caption{
Results with $ma = 0.024$. Top panel shows the chiral condensate for up quark and middle panel shows its susceptibility.
Bottom one is the binder ratio $B_4$.
\label{fig:result_ma0024} }
\end{figure}

\paragraph{Light mass regime}
Here we show results for $ma = 0.024$ and $ma = 0.028$.
The former belongs to the first order regime for $N_b=0$ \cite{Smith:2011pm} and results are summarized in \fig\ref{fig:result_ma0024} for the up quark.
Down and strange quarks show similar behavior except for the absolute magnitude.
The critical $\beta$ shifts as a function of $N_b$ monotonically,
showing the magnetic catalysis instead of the inverse one.

\begin{figure}[htbp]
        \begin{center}
              \begin{minipage}{0.55\hsize }
                \begin{center}
                        \includegraphics[width=\hsize]{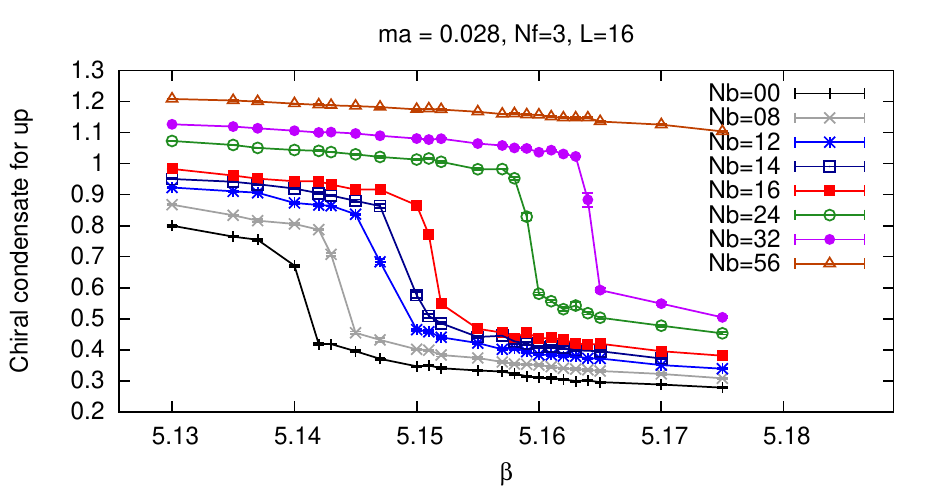}
                \end{center}
              \end{minipage}
              \begin{minipage}{0.55\hsize }
                \begin{center}
                        \includegraphics[width=\hsize]{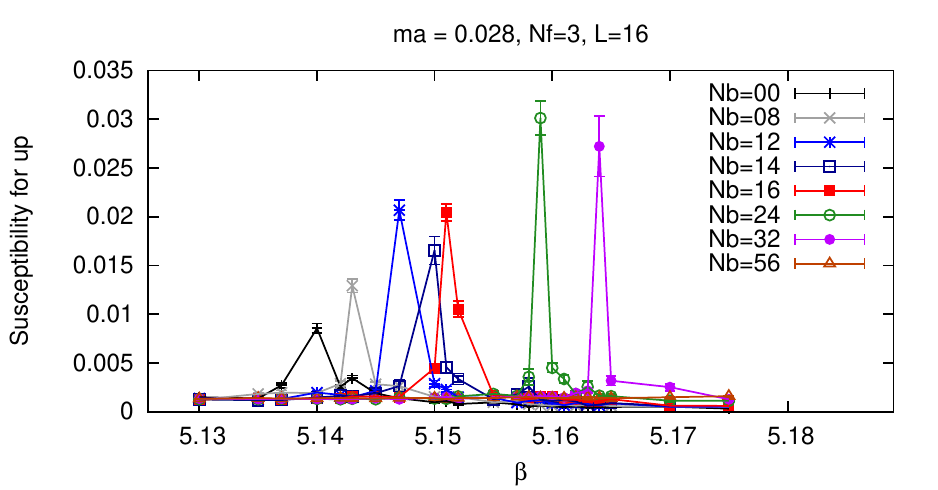}
                \end{center}
              \end{minipage}
                          \begin{minipage}{0.55\hsize }
                \begin{center}
                        \includegraphics[width=\hsize]{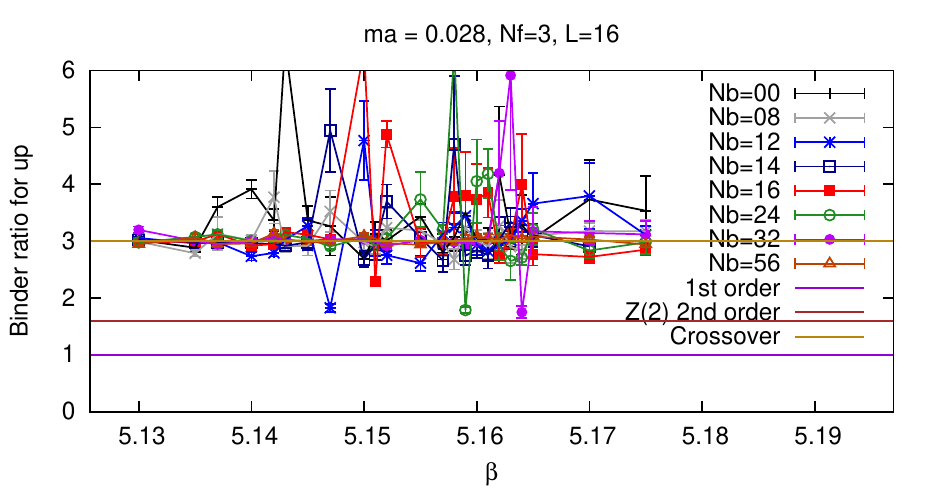}
                \end{center}
              \end{minipage}
        \end{center}
\caption{
Same plots of \fig \ref{fig:result_ma0024} but for $ma = 0.028$. 
\label{fig:result_ma0028} }
\end{figure}

The latter parameter corresponds to to the regime just above the critical mass for $N_b=0$
and the results are summarized in \fig\ref{fig:result_ma0028}.
Standard fermions with $ma=0.028$, and $N_\tau=4$ case, it shows a crossover-like behavior.
This is consistent with the Binder cumulant (Bottom of \fig\ref{fig:result_ma0028}, black line).
The $B$ dependence of $\beta_\text{crit}$ shows similar tendencies in the case of $ma = 0.024$.

The first order phase transition tends to become
stronger as magnetic field increases in contrast to a model prediction in \cite{mueller2015magnetic}.
We can see from the chiral condensate that the critical temperature increases with the magnetic field.

\paragraph{Heavy mass regime}
Here we show results with $ma = 0.2$, $ma = 0.4$ and $ma = 0.8$.
In all cases here chiral condensates do not show any critical behavior as expected.

\begin{figure}[htbp]
        \begin{center}
            \begin{tabular}{c}
              \begin{minipage}{0.5\hsize }
                \begin{center}
                        \includegraphics[width=\hsize]{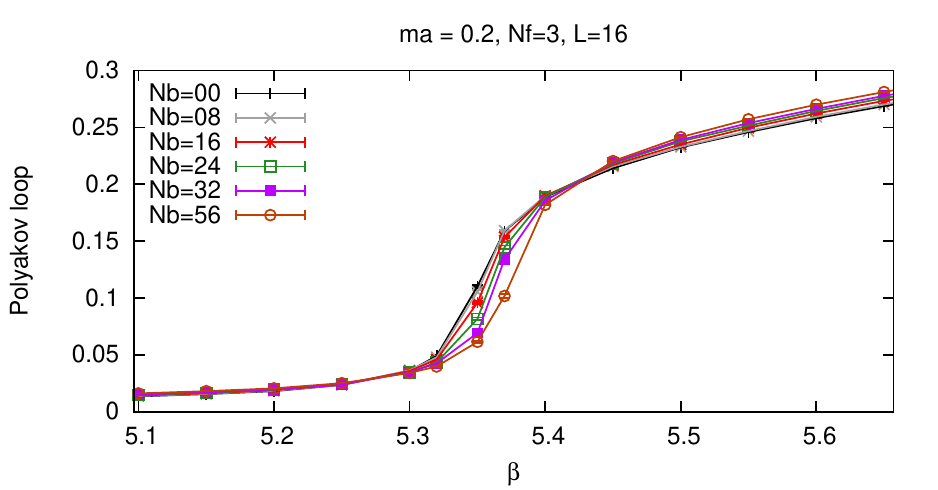}
                \end{center}
              \end{minipage}
              \begin{minipage}{0.5\hsize }
                \begin{center}
                        \includegraphics[width=\hsize]{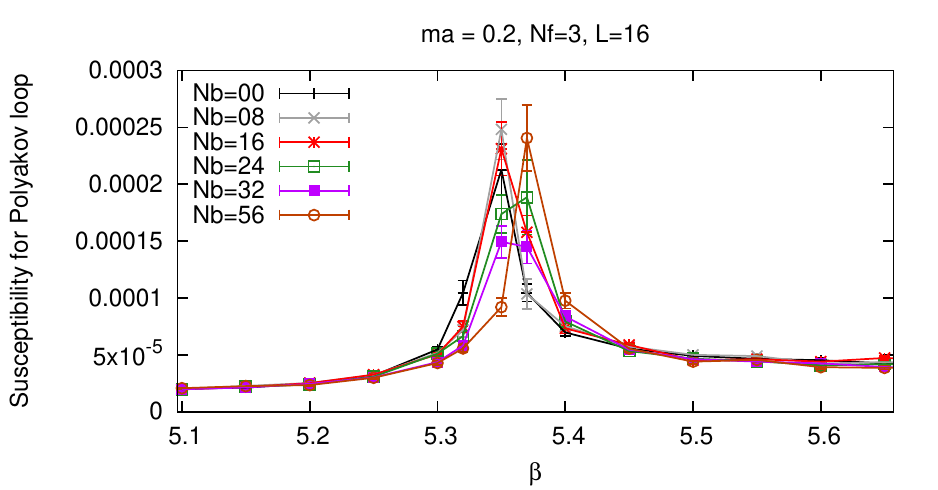}
                \end{center}
              \end{minipage}
            \end{tabular}
        \end{center}
\caption{
Results $ma = 0.2$. Left panels shows the Polyakov loop and right panel shows its susceptibility.
\label{fig:result_ma0200} }
\end{figure}

\begin{figure}[htbp]
        \begin{center}
            \begin{tabular}{c}
              \begin{minipage}{0.5\hsize }
                \begin{center}
                        \includegraphics[width=\hsize]{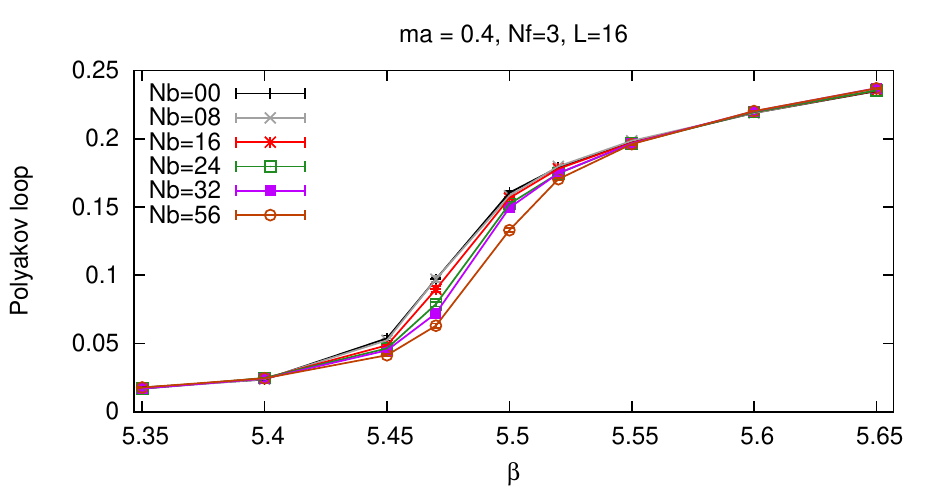}
                \end{center}
              \end{minipage}
              \begin{minipage}{0.5\hsize }
                \begin{center}
                        \includegraphics[width=\hsize]{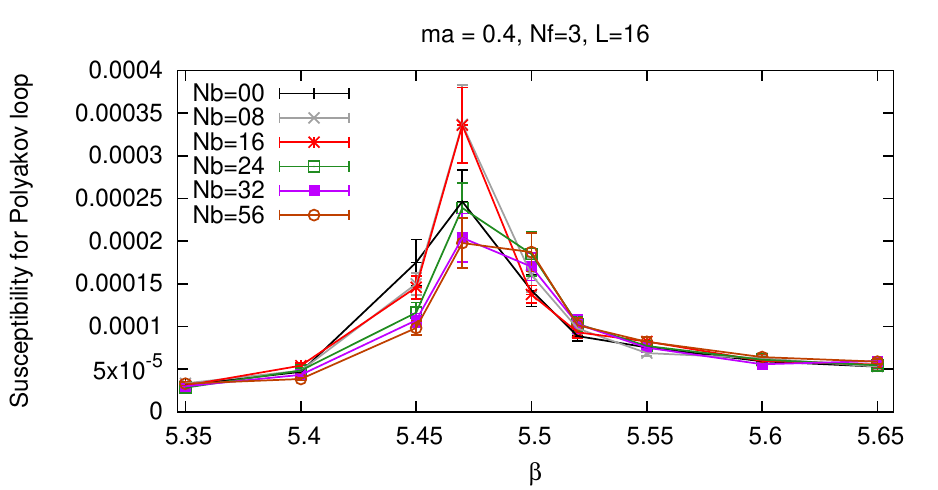}
                \end{center}
              \end{minipage}
            \end{tabular}
        \end{center}
\caption{
Same plots of \fig \ref{fig:result_ma0200} but for $ma = 0.4$. 
\label{fig:result_ma0400} }
\end{figure}

\begin{figure}[htbp]
        \begin{center}
            \begin{tabular}{c}
              \begin{minipage}{0.5\hsize }
                \begin{center}
                        \includegraphics[width=\hsize]{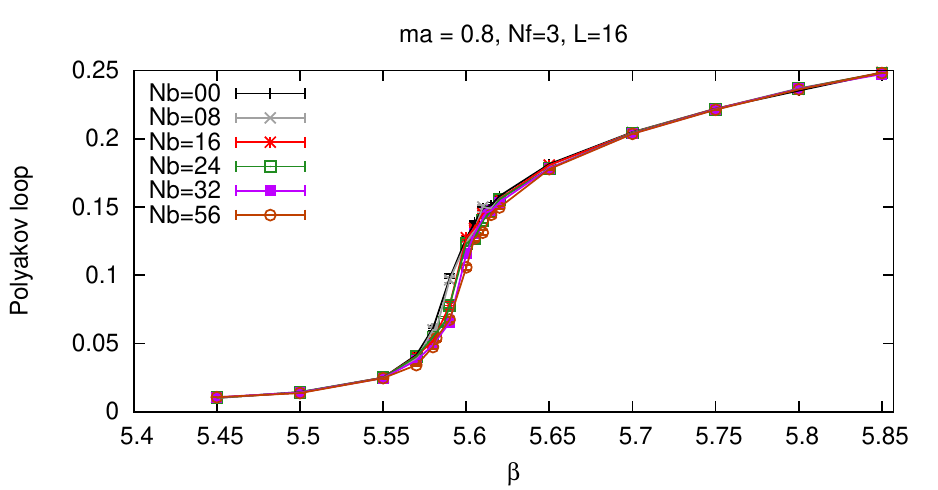}
                \end{center}
              \end{minipage}
              \begin{minipage}{0.5\hsize }
                \begin{center}
                        \includegraphics[width=\hsize]{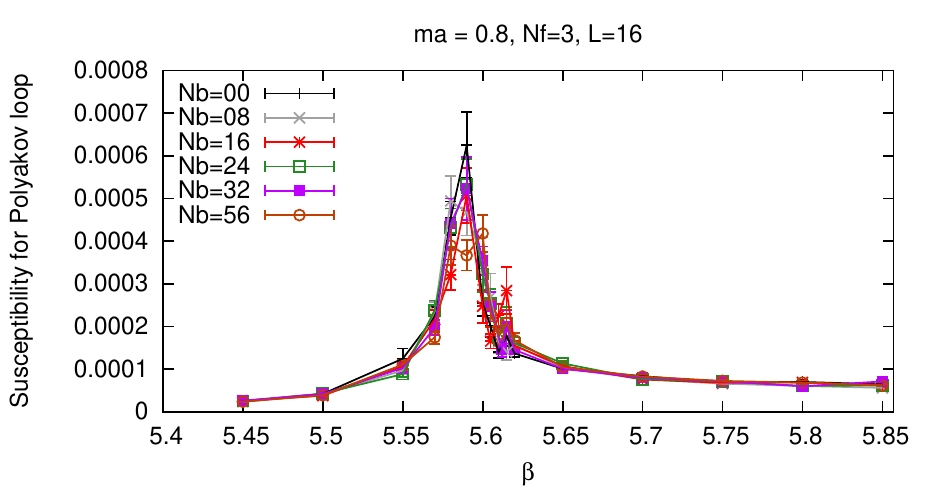}
                \end{center}
              \end{minipage}
            \end{tabular}
        \end{center}
\caption{
Same plots of \fig \ref{fig:result_ma0200} but for $ma = 0.8$. 
\label{fig:result_ma0800} }
\end{figure}

In \fig\ref{fig:result_ma0200} and \fig\ref{fig:result_ma0400} we show results of the Polyakov loop and its susceptibility for
$ma = 0.2$ and $ma = 0.4$ , respectively.
The confinement transition is slightly affected by the magnetic field. 
This is natural because gluons do not couple to the external magnetic field directly
and effects of the magnetic field here come through dynamical but very heavy quarks (almost quenched).
\fig\ref{fig:result_ma0800} shows results of the Polyakov loop and its susceptibility for $ma = 0.8$.
For this mass, effects of the magnetic field are tiny.
In this mass regime, the response of the Polyakov loop is similar to predicted by the PNJL model \cite{Fukushima:2010fe}.
In addition the magnetic field increases the critical temperature in the same way as in the lighter mass regime except for the heaviest mass case.

By comparing the values of the threshold mass and the quark mass it seems that in the current simulation
the QCD thermodynamics is more dominated by effects introduced by the quark mass.


\section{Summary and discussion}
\label{sec:summary}
In this work, we investigate the phase structure of QCD with three degenerate flavor and a \uone external magnetic field for various masses
using standard staggered fermions.
%
We have observed a tendency of strengthening of the phase transition in the light mass regime.
Except for the heaviest case,  the critical temperature increases with the magnetic field instead of decrease.

%
There are several issues which must be addressed in our forthcoming study.
Firstly, we need to increase the statistics and improve resolution of $\beta$
to obtain more accurate results especially the determination of the order of phase transition from the Binder cumulant.
Secondly, we find the increase of $\beta_\text{crit}$ instead of decrease.
That might come from cutoff effects (lattice artifacts) of our setup.
This is because, for the $N_\tau=4$ case,  the pion mass is relatively heavy (around 290 MeV \cite{Karsch:2001nf,Karsch:2003va}),
even in lighter quark mass regime.
We have used parameters as in \cite{Smith:2011pm}, which correspond to a first order chiral phase transition regime.
However, the existence of a first order phase transition regime for massless three flavor QCD is not clear yet (see  \cite{Bazavov:2017xul}
and the references therein).
In order to confirm or refine our results, we need to improve our lattice action to reduce the root mean square  pion mass and lattice artifacts.
This is important because the chiral features strongly depends on the low-laying modes of the Dirac operator even in the case of vanishing magnetic field,
which are tightly connected to discretization of the Dirac operator.
Even more, low-laying modes with magnetic field, namely the lowest Landau level, plays essential role when magnetic field is applied \cite{kojo2014quark}.
Thus, we have to use an improved action in the next step and clarify which phenomena are coming from continuum physics and 
which are just lattice artifacts.

\section*{Acknowledgement }
We thank M. D’Elia for giving us exact numbers of their original work.
AT thank the members of HotQCD collaboration for fruitful discussions and Y. Nakamura for providing information of current status for three flavor massless QCD in the critical regime.
The simulations are performed at {\it Fermilab Lattice Gauge Theory Computational Facility}.
The work of AT was supported in part by NSFC under grant no. 11535012.

\bibliography{bib}

\end{document}